# Randomization Using Quasigroups, Hadamard and Number Theoretic Transforms

V. Spoorthy Ella

*Abstract.* This paper investigates the use of quasigroups, Hadamard transforms and Number Theoretic Transforms for use in sequence randomization. This can also be used to generate hash functions for sequence encryption.

## I INTRODUCTION

Good symmetric encryption schemes as well as randomization and hashing techniques are based on effective techniques of confusion and diffusion [1]. Quasigroups provide an excellent way to generate an astronomical number of keys and therefore they are excellent at confusion [2] but they are not equally good at diffusing the statistics of the plaintext. Specifically, the quasigroup transformation can be easily discovered by the known plaintext attack. For quasigroup mappings in encryption, it is necessary, therefore, to use this mapping together with other statistics-diffusing mappings.

A quasigroup ($Q$, *) is a set of numbers with a binary operation * such that for each *a* and *b* in *Q*, there exist unique elements *x* and *y* in *Q* such that:

$a*x = b$
$y*a = b$

In other words, for two elements *a* and *b*, *b* can be found in row *a* and *a* in column *b* of the quasigroup table of element operations. The group operations thus are equivalent to a table of permutations.

**Example**. A quasigroup of order 7, which consists of elements 0,1,2,3,4,5,6:

**Index**

| * | 0 | 1 | 2 | 3 | 4 | 5 | 6 |
|---|---|---|---|---|---|---|---|
| **0** | 2 | 1 | 0 | 5 | 4 | 6 | 3 |
| **1** | 1 | 4 | 3 | 2 | 0 | 6 | 5 |
| **2** | 0 | 5 | 1 | 6 | 3 | 4 | 2 |
| **3** | 4 | 3 | 6 | 1 | 2 | 5 | 0 |
| **4** | 6 | 2 | 5 | 0 | 1 | 3 | 4 |
| **5** | 3 | 0 | 2 | 4 | 5 | 1 | 6 |
| **6** | 5 | 6 | 4 | 3 | 0 | 2 | 1 |

Suppose *a*=6 and *x*=3 and, then in $a*x = b$, *b*=3.

**Encryption using quasigroup:** Consider ($a_1, a_2, a_3...a_n$) are input elements; $a$ is the seed element, and ($e_1, e_2, e_3...e_n$) are output elements after the quasigroup transformation.

$e_1=a*a_1$
$e_2=e_1*a_2$
$e_3=e_2*a_3$
.
.
.
$e_n=e_{n-1}*a_n$

**EXAMPLE:** Seed element a=3

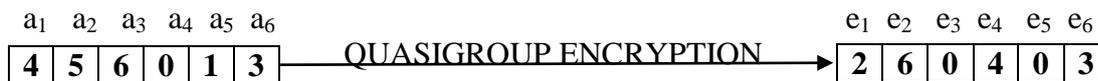

$a_1\ a_2\ a_3\ a_4\ a_5\ a_6$ → QUASIGROUP ENCRYPTION → $e_1\ e_2\ e_3\ e_4\ e_5\ e_6$
| 4 | 5 | 6 | 0 | 1 | 3 | → | 2 | 6 | 0 | 4 | 0 | 3 |

$e_1=a*a_1=3*4=2$
$e_2=e_1*a_2=0*5=6$
$e_3=e_2*a_3=3*6=0$
$e_4=e_3*a_4=3*0=4$
$e_5=e_4*a_5=5*1=0$
$e_6=e_5*a_6=6*3=3$

One way to diffuse statistics effectively is the use of transforms [3] where the security is enhanced by using a variety of them and by chaining them [4]. Here we investigate the use of chained Hadamard transforms and NTTs (number theoretic transforms) to introduce diffusion together with the quasigroup transformation.

We can overcome these drawbacks by performing additional transformations along with quasigroup scrambling. The Hadamard transform is a generalized class of discrete Fourier transforms [4]. It is generated either recursively, or by using binary representation. All the values in the matrix are non-negative. Each negative number is replaced with corresponding modulo number. For example in modulo 7 Hadamard matrixes -1 is replaced with 6 to make the matrix non-binary. Due to its symmetric form it can be used in applications such as data encryption and randomness measures. Only prime modulo operations are performed because non-prime numbers can be divisible with numbers other than 1 and itself.

Number Theoretic Transforms are also a type of discrete Fourier transforms. The N*umber theoretic transform* is based on generalizing the nth primitive root of unity to a quotient ring instead of using complex numbers.

In this paper, the input sequence will undergo different transformations sequentially like quasigroup transformation, Hadamard transformation and Number theoretic transformation. For Hadamard and Number theoretic transforms input is split into certain group of bits such that each group bit count is the order of the corresponding matrix.

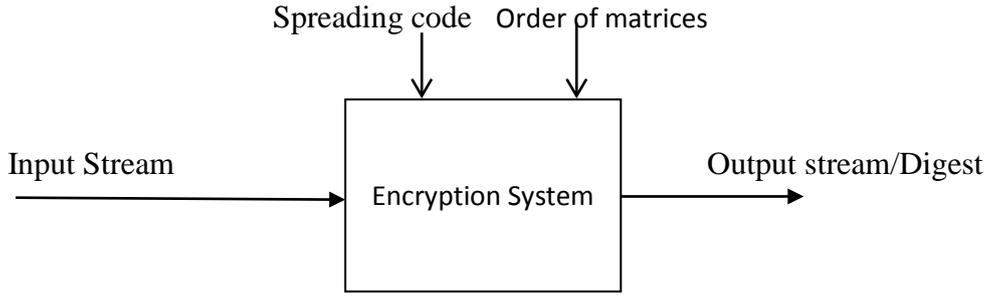

Figure 1 The General Architecture of the proposed encryption and hashing system

## II HADAMARD TRANSFORMS

The Hadamard transform is a generalized class of discrete Fourier transforms [4],[18],[19]. It is generated either recursively, or by using binary representation. All the values in the matrix are non-negative. Each negative number is replaced with corresponding modulo number. For example in modulo 7 Hadamard matrixes -1 is replaced with 6 to make the matrix non-binary. Due to its symmetric form it can be used in applications such as data encryption and randomness measures [5]-[8]. Only prime modulo operations are performed because non-prime numbers can be divisible with numbers other than 1 and itself [2]. Recursively, we define the $1 \times 1$ Hadamard transform $H_0$ by the identity $H_0 = 1$, and then define $H_m$ for $m > 0$ by,

$$H_m = \frac{1}{\sqrt{2}} \begin{pmatrix} H_{m-1} & H_{m-1} \\ H_{m-1} & -H_{m-1} \end{pmatrix}$$

A Hadamard matrix, $\mathbf{H}_n$, is a square matrix of order n = 1, 2 or 4k where k is a positive integer. The elements of $\mathbf{H}$ are either +1 or –1 and $\mathbf{H}_n \cdot \mathbf{H}_n^T = n\mathbf{I}_n$, where $\mathbf{H}_n^T$ is the transpose of $\mathbf{H}_n$, and $\mathbf{I}_n$ is the identity matrix of order n. A Hadamard matrix is said to be normalized if all of the elements of the first row and first column are +1. Some examples of the Hadamard matrices are given below.

$$H_{o} = +1$$
$$H_1 = \frac{1}{\sqrt{2}} \begin{pmatrix} 1 & 1 \\ 1 & -1 \end{pmatrix}$$

Hadamard matrix of modulo 7 of size 8*8

| 1 | 1 | 1 | 1 | 1 | 1 | 1 | 1 |
|---|---|---|---|---|---|---|---|
| 1 | 6 | 1 | 6 | 1 | 6 | 1 | 6 |
| 1 | 1 | 6 | 6 | 1 | 1 | 6 | 6 |
| 1 | 6 | 6 | 1 | 1 | 6 | 6 | 1 |
| 1 | 1 | 1 | 1 | 6 | 6 | 6 | 6 |
| 1 | 6 | 1 | 6 | 6 | 1 | 6 | 1 |
| 1 | 1 | 6 | 6 | 6 | 6 | 1 | 1 |
| 1 | 6 | 6 | 1 | 6 | 1 | 1 | 6 |

Hadamard matrix of modulo 31 of size 8*8

| 1 | 1  | 1  | 1  | 1  | 1  | 1  | 1  |
|---|----|----|----|----|----|----|----|
| 1 | 30 | 1  | 30 | 1  | 30 | 1  | 30 |
| 1 | 1  | 30 | 30 | 1  | 1  | 30 | 30 |
| 1 | 30 | 30 | 1  | 1  | 30 | 30 | 1  |
| 1 | 1  | 1  | 1  | 30 | 30 | 30 | 30 |
| 1 | 30 | 1  | 30 | 30 | 1  | 30 | 1  |
| 1 | 1  | 30 | 30 | 30 | 30 | 1  | 1  |
| 1 | 30 | 30 | 1  | 30 | 1  | 1  | 30 |

Hadmard matrix of modulo 7 of size 4*4

| 1 | 1 | 1 | 1 |
|---|---|---|---|
| 1 | 6 | 1 | 6 |
| 1 | 1 | 6 | 6 |
| 1 | 6 | 6 | 1 |

The idea of encryption is to multiply the decimated input sequence with the non-binary Hadamard matrix in a chained manner block by block. The block size depends upon the size of the Hadamard matrix you have chosen. Input sequence is taken in the form of column matrix.

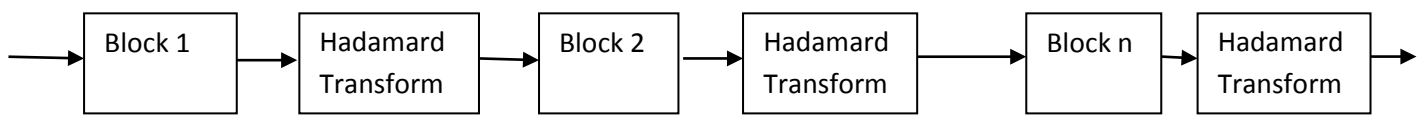

Figure 2 Hadamard encryption

### III NUMBER THEORETIC TRANSFORMS

Number theoretic transforms are also a type of discrete Fourier transforms. The N*umber theoretic transform* is based on generalizing the nth primitive root of unity to a quotient ring instead of using complex numbers [9].

$$\begin{pmatrix} 1 & 1 & 1 & 1 \\ 1 & w & w^2 & w^3 \\ 1 & w^2 & w^4 & w^6 \\ 1 & w^3 & w^6 & w^9 \end{pmatrix}$$

The unit w is exp(2π/ n). Now everything a number theoretic transform is all about is that $w^n=1$. We do Number theoretic transform it in matrix form by multiplying your data with a Fourier matrix.

**NTT matrix of order 6*6**

| 1 | 1 | 1 | 1 | 1 | 1 |
|---|---|---|---|---|---|
| 1 | 3 | 2 | 6 | 4 | 5 |
| 1 | 2 | 4 | 1 | 2 | 4 |
| 1 | 6 | 1 | 6 | 1 | 6 |
| 1 | 4 | 2 | 1 | 4 | 2 |
| 1 | 5 | 4 | 6 | 2 | 3 |

**NTT Encryption:**

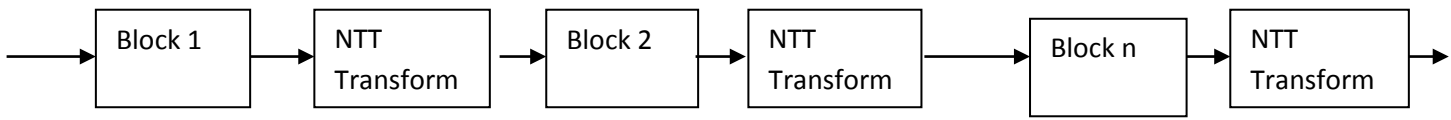

Figure 3  Number Theoretic encryption

The idea of encryption is to multiply the decimated input sequence (which is the output we have obtained after encryption using Hadamard transform with the non-binary Number theoretical matrix in a chained manner block by block. The block size depends upon the size of the Number theoretical matrix you have chosen. Input sequence is taken in the form of column matrix.

## IV THE ENCRYPTION SYSTEM

**Encryption:**

**Phase1:** Encryption of input data using quasigroup based encryption system.
**Phase2:** Output of Phase1 is given as input to the Phase 2.In phase2 we perform Hadmard transformation of data by multiplying the data with Hadamard matrix.
**Phase3:** Output of Phase2 is given as input to the Phase 3.In phase 3 we perform Number Theoretic transform by multiplying the data with Number Theoretic Transform.
**Phase4:** Phase2 is repeated with a different order of Hadamard matrix.

**Decryption:**

As the Hadamard matrix operations are invertible, we can perform decryption of the data by generating inverse Hadamard matrix

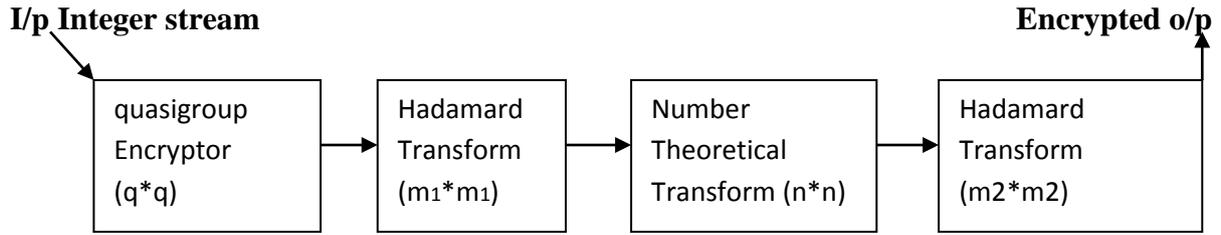

Figure 4 The Encryption system

1. All the matrices quasigroup, Hadamard Matrix and Number Theoretic transform matrix should have different orders so that at each encryption level the size of the input block size differs which eventually increases the randomness of the input sequence.
2. Hadamard transforms and Number Theoretic transforms work as hash functions which generate different hash values for different input values [2].
3. For other functions like triangular function, input consisting of all zeros and consisting of all ones it generates a Pseudo random sequence.
4. If one bit changes in the input sequence, there is a large difference in the generated random sequence.
5. Input size should be multiple of orders of all three matrices i.e. quasigroup, Hadamard and Number theoretic matrix in order to assume the block size.

## V PERFORMANCE MEASURE OF THE SYSTEM

**Method 1:** Using Autocorrelation Function.

**Tests of Randomness:**

We use autocorrelation functions to test the randomness of input and output data. If the given sequence is random, autocorrelation values should be nearer to zero.

**Function of Autocorrelation:**

$$C(k) = \frac{1}{n}\sum_{j=1}^{n}(a_j a_{j+k})$$

Where n is period and k=0 to n-1.

**Randomness Measure:** Randomness R of a sequence of period n is measured by the following formula. If the given sequence is random, R value should be nearer to 1.

$$R(Sequence) = 1 - \frac{1}{n-1}\sum_{k=1}^{n-1}(|C(k)|)$$

While calculating autocorrelation of our encrypted system we replace all integers with their corresponding binary values 0s and 1s. And 0s are replaced with -1.

**Example 1:** Here Input size=684, Period=3*684=2052. (We are converting the input integer to 3 bit binary form)

Characteristics of input data**:** Pseudo random stream generates by java random function.

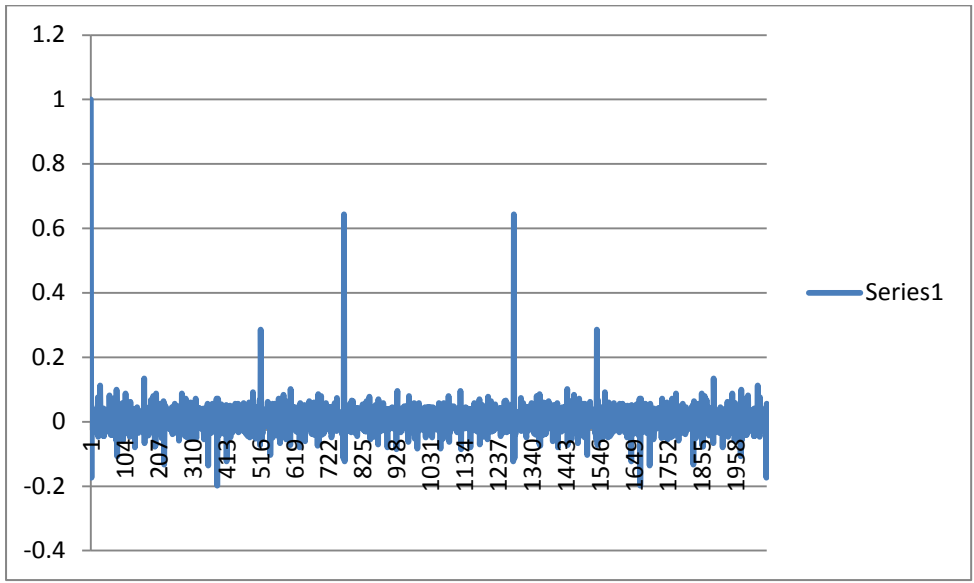

Figure 5 Autocorrelation graph for input Pseudo random sequence and R=0.9428.

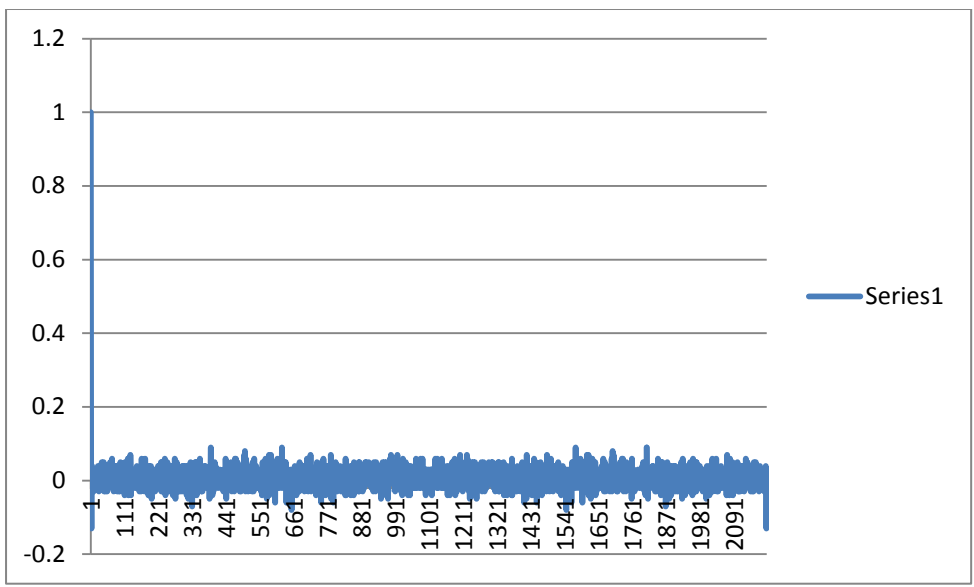

Figure 6 Autocorrelation graph of the output sequence generated by the encryption system R=0.9631

**Example 2:** Input size=3200, Period of autocorrelation=9600

Characteristics of input data**:** Pseudo random stream generates by java random function.

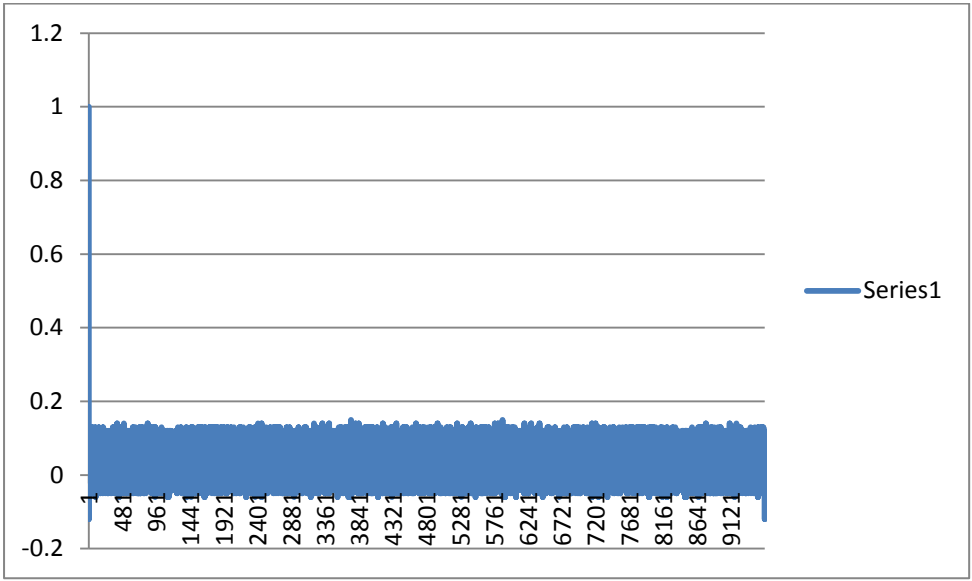

Figure 7  Autocorrelation graph for input Pseudo random sequence R=0.9538

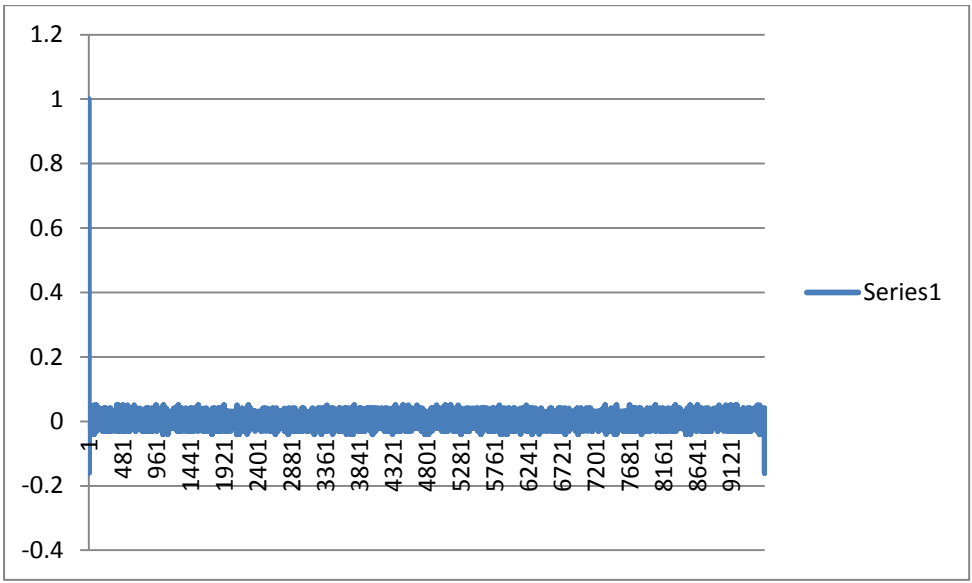

Figure 8  Autocorrelation graph of the output sequence generated by the encryption system. R=0.9724

**Example 3:** Input type: d-sequences [11]-[16].

Characteristics of d-sequence: prime number: 2029, period: 2028

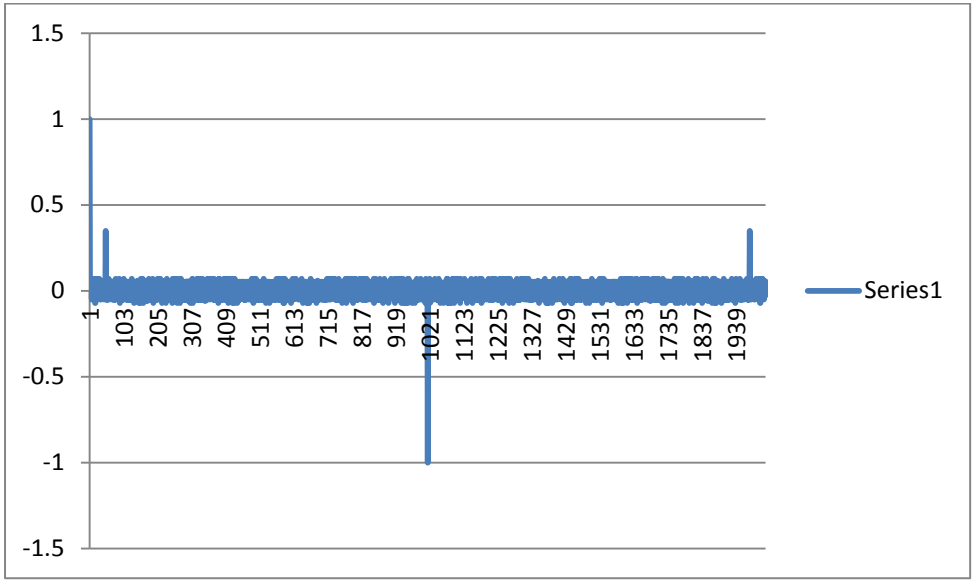

Figure 9 Autocorrelation graph for input d-sequence and R=0.9238

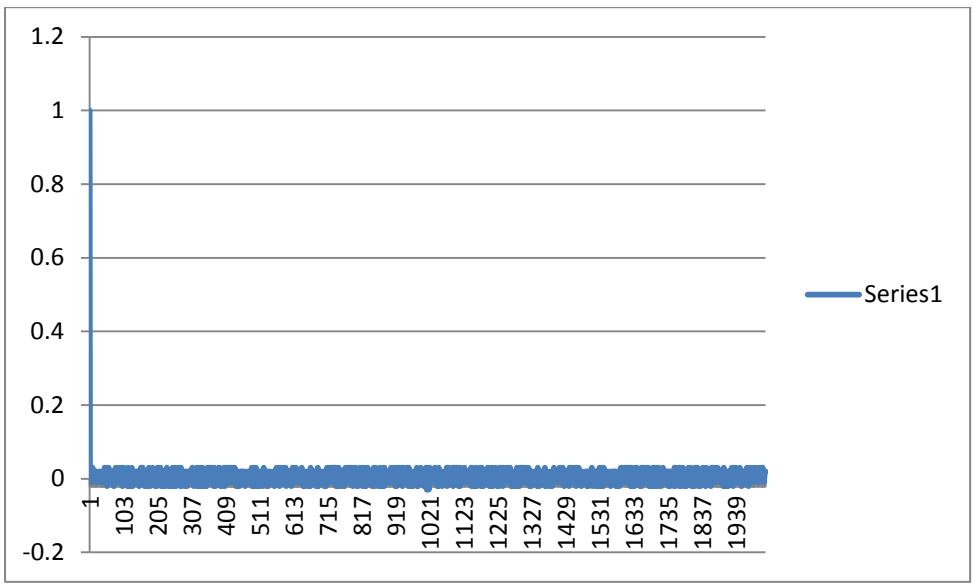

Figure 10 Autocorrelation graph for output d-sequence and R=0.9798

**Example 4:** Input size=684, period=3*684=2052

Characteristics of Input data: all zeros and last bit as '1'.

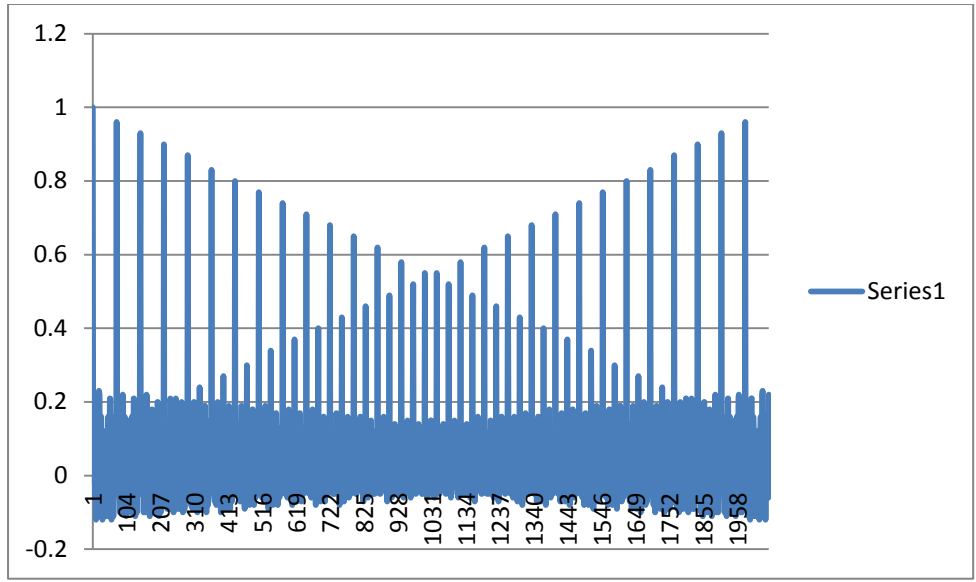

Figure 9  Autocorrelation graph of input sequence with all zeros and last bit as 1. R=0.9072

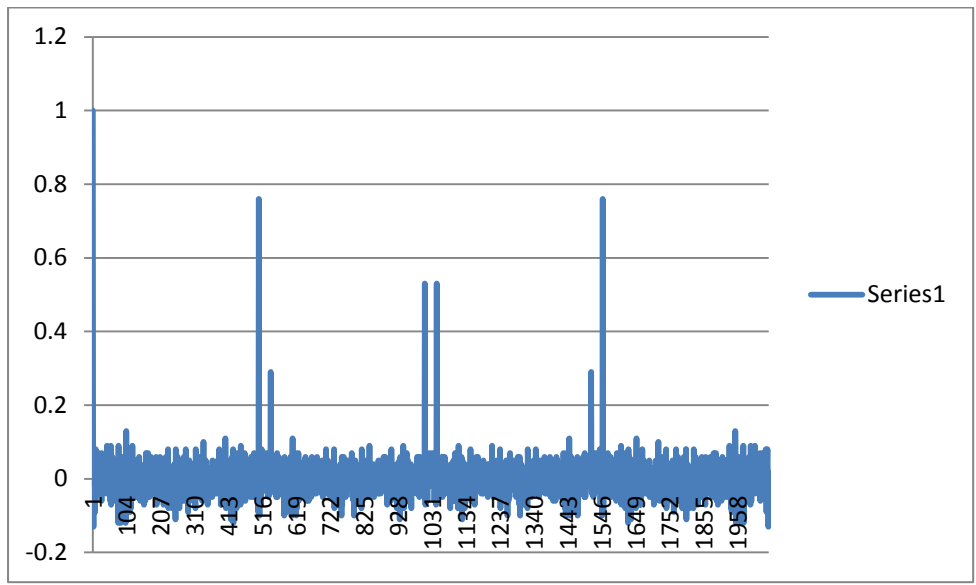

Figure 10 Autocorrelation graph of output sequence. R=0.9515

In this case we get output as Pseudo random sequence because we have taken the input with all zeros and only last bit is 1, quasigroup elements are repeated and Hadamard and Number Theoretic Transforms has no effect. We can get better results if we consider quasigroup, Hadamard and Number theoretic matrix having different orders as it is same as the block size of the transformation at each phase.

**Example 5:** Input size: 684, period: 2052

Characteristics of Input Data: all 1s

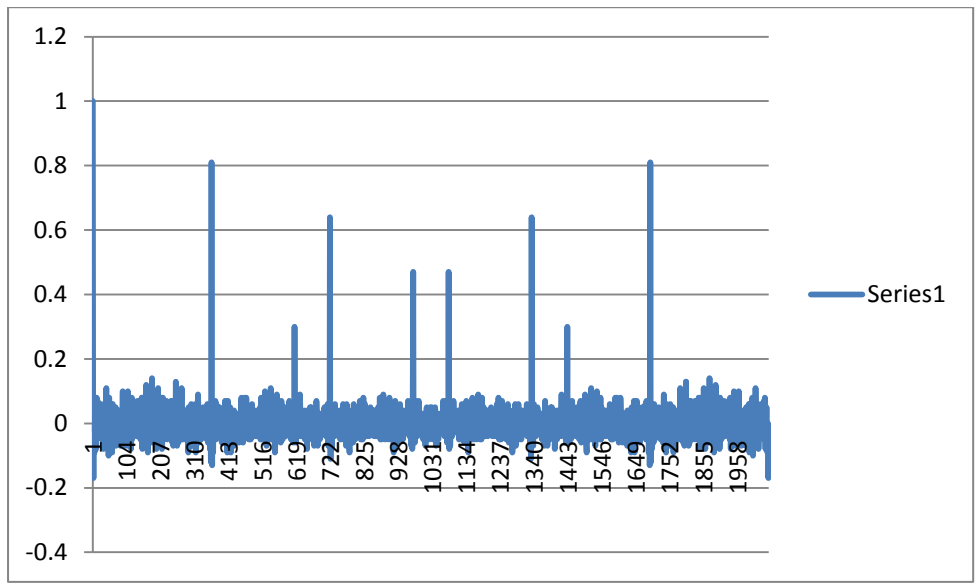

Figure 11 Autocorrelation graph of output sequence where the input sequence contains all 1s
R=0.9521

In this case, the input sequence consists of all ones. The output sequence contains Pseudo random data. At some points the auto correlation values are high which is due to the input size we have taken and block sizes which are nothing but order of the matrices Quasigroup, Hadamard and Number theoretic matrix).

**Method 2:** Using Chi Square Statistics.

Frequency test within a block: The purpose of this test is find proportion of ones within M-bit blocks. According to this statistics the frequency of ones in an m-bit block is approximately

M/2 would be expected under an assumption of randomness.

**Parameters:**

M=the length of each block.

n=the length of the bit string.

$\chi^2_{(obs)=}$A measure of how well the observed proportion of ones within a given M-bit block match the expected proportion(1/2).

The reference distribution for the test statistic is a $\chi^2$ distribution.

ε= the sequence of bits as generated by the random or Pseudo random being tested.

**Test procedure:**

**Step 1:** Partition the input sequence into N=⌊n/M⌋ non-overlapping blocks. Discard any unused bits.

**Step 2:** Determine the proportion $\pi_i$ of ones in each M-bit block using the equation $\pi_i = \frac{\sum_{j=1}^{M} \varepsilon_{(i-1)M+j}}{M}$ for $1 \leq i \leq N$.

**Step 3:** Compute the $\chi^2$ statistic: $\chi^2(obs) = 4M \sum_{i=1}^{N} (\pi_i - \frac{1}{2})^2$.

**Step 4:** Compute P-value =**igamc** (N/2, $\chi^2$(obs)/2), where **igamc** is the incomplete gamma function.

**Decision Rule:** (at the 1% Level)

If the computed P-value is <0.01, then conclude that the sequence is non-random. Otherwise, conclude that the sequence is random.

**Example 6:** Sequence: Output generated using proposed encryption system where input is a pseudo random sequence generated by the Java random function.

Non binary sequence size: 684

Binary sequence size=684*3=2052

Block size (M)=18 hence Number of blocks (N)=2052/18=114.

We get $\chi^2$ value as 58.

P-value=igamc(114/2,58)=0.56977 >0.01, hence the generated sequence is random.

**Example 7:** Sequence: Output generated using proposed encryption system where input is a pseudo random sequence generated by the Java random function.

Non binary sequence size: 636

Binary sequence size=636*3=1908

Block size (M) =18 hence Number of blocks (N) =1908/18=106.

We get $\chi^2$ value as 52.

P-value=igamc(106/2,52)=0.46320 >0.01, hence the generated sequence is random.

## VI CONCLUSION

We have shown that the encryption system which consists of sequential transforms like quasigroup scrambling, Hadamard transform and Number Theoretic transform provides a method of generating random output sequence and also can act as hash function generator. The randomization obtained is very good, and, therefore, one can foresee practical applications for this method.

## REFERENCES

[1] C.E. Shannon, Communication theory of secrecy systems. Bell System Technical Journal 28: 656-715, 1949.

[2] M. Satti and S. Kak, Multilevel indexed quasigroup encryption for data and speech. IEEE Trans on Broadcasting 55: 270-281, 2009.

[3] J. Hoffstein, J. Pipher, J.H. Silverman, An Introduction to Mathematical Cryptography. Springer, 2010.

[2] R.S. Reddy, Encryption of binary and non-binary data using chained Hadamard transforms. arXiv:1012.4452

[3] V.Godavarty, Using Quasigroups for Generating Pseudorandom Numbers. arXiv:1112.1048

[4] B. Goldburg, S. Sridharan, E. Dawson, Design and cryptanalysis of transform-based analog speech scramblers. Journal of Selected Areas in Communications 11: 735-744, 1993.

[5] V.Milosevic, V.Delic, V.Senk, Hadamard transform application in speech scrambling. Proceedings 13$^{th}$ Intl. Conference on Digital Signal Processing, 361-364, 1997

[6] C. C. Gumas, A century old fast Hadamard transform proves useful in digital communications, Chip Centerquestlink, 2006.

[7] S. Kak, Classification of random binary sequences using Walsh-Fourier analysis. IEEE Trans. on Electromagnetic Compatibility EMC-13, pp. 74-77, 1971.

[8] S. Kak, Joint encryption and error-correction coding, Proceedings of the 1983 IEEE Symposium on Security and Privacy, pp. 55-60 1983.

[9] T. Koshy, Elementary Number Theory with Applications, 2nd Ed, Elsevier Inc., Academic Press Publications, Burlington, MA, 2007, pp. 346.

[10] A. Rukhin et al, A Statistical Test Suite for Random and Pseudorandom Number Generators for Cryptographic Applications. National Institute of Standards and Technology Gaithersburg, April 2010.

[11]     S. Golomb, Shift Register Sequences. San Francisco, Holden–Day, 1967

[12]     S. Kak and A. Chatterjee, On decimal sequences. IEEE Transactions on Information Theory, IT-27: 647 – 652, 1981.

[13]     S. Kak, Encryption and error-correction coding using D sequences. IEEE Transactions on Computers C-34: 803-809, 1985.

[14]     S. Kak, New results on d-sequences. Electronics Letters 23: 617, 1987.

[15]     D. Mandelbaum, On subsequences of arithmetic sequences. IEEE Trans on Computers 37: 1314-1315, 1988.

[16]     S Kak, Prime reciprocal digit frequencies and the Euler zeta function. arXiv:0903.3904

[17] R. Kandregula, The basic discrete Hilbert transform with an information hiding application. arXiv:0907.4176v1

[18] S. Kak, Hilbert transformation for discrete data. International Journal of Electronics 34: 177-183, 1973.

[19] S. Kak, The discrete finite Hilbert transform. Indian Journal of Pure and Applied Mathematics 8: 1385-1390, 1977.